\documentclass{sigchi}

\usepackage{balance}       
\usepackage{graphics}      
\usepackage[T1]{fontenc}   
\usepackage{txfonts}
\usepackage{mathptmx}
\usepackage[pdflang={en-US},pdftex]{hyperref}
\usepackage{color}
\usepackage{booktabs}
\usepackage{textcomp}
\usepackage{csquotes}
 \usepackage{cite}
\usepackage{microtype}        
\usepackage{ccicons}          

\usepackage{todonotes}

\newcommand{\sys}{ScreenTrack}
\newcommand{\capsys}{SCREENTRACK}

\newcommand{\out}[1]{{#1}}

\newcommand{\TODO}[1]{\out{{\small\textcolor{red}{\bf [TODO: #1]}}}}

\newcommand{\revise}[1]{\out{{#1}}}

\def\plaintitle{\sys{}: Using a Visual History of a Computer Screen to Retrieve Documents and Web Pages}

\def\emptyauthor{}
\def\plainkeywords{Productivity; Task resumption; Self-tracking.}

\makeatletter
\def\url@leostyle{%
  \@ifundefined{selectfont}{
    \def\UrlFont{\sf}
  }{
    \def\UrlFont{\small\bf\ttfamily}
  }}
\makeatother
\def\pprw{8.5in}
\def\pprh{11in}

\definecolor{linkColor}{RGB}{6,125,233}
\hypersetup{%
  pdftitle={\plaintitle},
  pdfauthor={\emptyauthor},
  pdfkeywords={\plainkeywords},
  pdfdisplaydoctitle=true, 
  bookmarksnumbered,
  pdfstartview={FitH},
  colorlinks,
  citecolor=black,
  filecolor=black,
  linkcolor=black,
  urlcolor=linkColor,
  breaklinks=true,
  hypertexnames=false
}

\toappear{\scriptsize Permission to make digital or hard copies of all or part of this work for personal or classroom use is granted without fee provided that copies are not made or distributed for profit or commercial advantage and that copies bear this notice and the full citation on the first page. Copyrights for components of this work owned by others than the author(s) must be honored. Abstracting with credit is permitted. To copy otherwise, or republish, to post on servers or to redistribute to lists, requires prior specific permission and/or a fee. Request permissions from permissions@acm.org. \\
{\emph{CHI '20, April 25--30, 2020, Honolulu, HI, USA.} } \\
\copyright~2020 Association for Computing Machinery. \\
ACM ISBN 978-1-4503-6708-0/20/04\ ...\$15.00. \\
http://dx.doi.org/10.1145/3313831.3376753}

\begin{document}

\title{\plaintitle}

\numberofauthors{2}
\author{
  \alignauthor{Donghan Hu\\ 
    \affaddr{Computer Science\\Center for Human-computer Interaction(CHCI)\\ Virginia Tech.}\\
    \email{hudh0827@vt.edu}
    }\\
    \alignauthor{Sang Won Lee\\ 
    \affaddr{Computer Science\\Center for Human-computer Interaction(CHCI)\\ Virginia Tech. }\\
    \email{sangwonlee@vt.edu}
    }\\
}\maketitle
\begin{abstract}
Computers are used for various purposes, so frequent context switching is inevitable. In this setting, retrieving the documents, files, and web pages that have been used for a task can be a challenge. While modern applications provide a history of recent documents for users to resume work, this is not sufficient to retrieve all the digital resources relevant to a given primary document. The histories currently available do not take into account the complex dependencies among resources across applications. To address this problem, we tested the idea of using a visual history of a computer screen to retrieve digital resources \revise{within a few days} of their use through the development of \sys{}. \sys{} is software that captures screenshots of a computer at regular intervals. It then generates a time-lapse video from the captured screenshots and lets users retrieve a \revise{recently} opened document or web page from a screenshot \revise{after recognizing the resource by its appearance}. A controlled user study found that participants were able to retrieve requested information more quickly with \sys{} than under the baseline condition with existing tools. A follow-up study showed that the participants used \sys{} to retrieve previously used resources and to recover the context for task resumption.
\end{abstract}

\vspace{-5pt}
\vspace{-5pt}

\section{Introduction}

It is common for users to use multiple applications (e.g., programming tools, office software, web browsers, calendars, communication apps) simultaneously when carrying out computing tasks~\cite{Bannon:1983:EAU:800045.801580}. 
However, users often have to switch from one task to another, and they may be interrupted before completing a task~\cite{Czerwinski:2004:DST:985692.985715,Gonzalez:2004:CCM:985692.985707}. 
Users therefore need to frequently retrieve digital resources --- e.g., software, files, and web pages --- that they used in the past~\cite{Dragunov:2005:TDE:1040830.1040855_tasktracer,Iqbal:2007:DRC:1240624.1240730}. 
This retrieval process can be time-consuming and cognitively challenging. 
A needed file may be located in a poorly organized folder structure, and the file name may not be informative (e.g., \texttt{/Users/janedoe/Desktop/00291294.pdf}). 
One common approach to this problem is to check the history of recently opened documents (or web pages) --- file names or URLs sorted in chronological order, for example --- in a particular application. 
However, there may be too many documents to choose from, and it may be difficult to recognize the correct item by its path, title, or URL alone. 
Furthermore, there may be other resources associated with a given primary document scattered across applications, and these are not connected to the document unless users manually organize such resources for each task they perform, which is challenging given the dynamic nature of modern knowledge work. 
Therefore, these additional resources can be easily lost in time or in multiple open windows.  
Eventually, users may need to rely on their memory, be forced to reopen and sort through a large number of resources, or fail to retrieve the resources entirely.
The overhead in task resumption can be greatly reduced if users can retrieve primary documents and relevant resources efficiently.


\begin{figure*}
\centering
  \includegraphics[width=\textwidth]{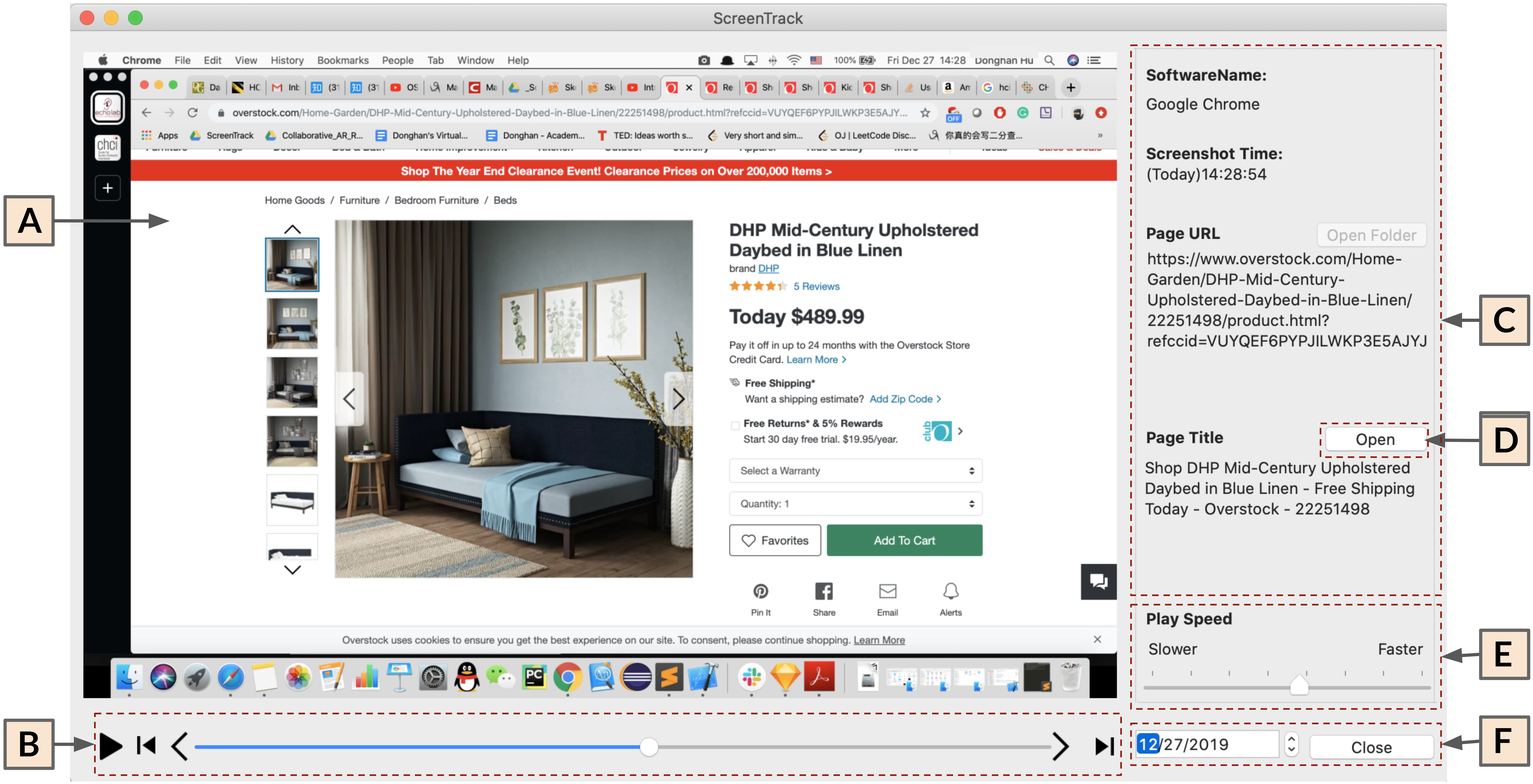}
  \caption{\sys{}: a desktop application that captures screenshots at regular intervals to let users retrieve previously used software, documents, and web pages from the screenshot. \sys{} leverages users' episodic memory and provides multimodal cues (visual and metadata) for users to recognize digital resources and retrieve them. For example, a user may not remember an author's name, the URL of a page, or \revise{the title of a web page that they visited, but they may be able to recognize an image they saw}. (A) A time-lapse video of computer screenshots; (B) UI controls for temporal navigation patterned after a video player; (C) metadata, such as application names, URLs, page titles, and filenames for the frontmost (active) application; (D) button to open retrieved items; (E) playback speed control; (F) date picker and close button}~\label{fig:userinterface}
  \vspace{-10pt}
\end{figure*}
To address this problem, we present a way to retrieve previously used digital resources --- software, websites, or files --- from a visual history of a computer screen.
We prototyped the method in \sys{}(Figure~\ref{fig:userinterface}), a desktop application which records screenshots and lets users open digital resources present in the screenshots.  
We draw ideas from previous works that use a series of computer screenshots, or thumbnails of applications, to help users reconstruct their mental context~\cite{rule:hal-01213708,Russell2014, Rule:2017:UVH:3149968.3149977, Russell_Retrospective,kaasten2002people}.
\sys{} takes screenshots regularly, at a user-specified interval (e.g., every 10 seconds) to record current computer activities. 
It captures any metadata available from the operating system, such as the frontmost application, the path to an open file, or the URL to an open web page, and associates that information with the screenshot. 
\sys{} then provides users with a time-lapse video made of captured screenshots for a limited period of time (N days) from which they can navigate the visual history and reopen closed software, documents, or URLs (see Figure~\ref{fig:userinterface}). 

\revise{\sys{} is particularly designed for a short-term scenario; ideally, a user knows the approximate time when a document was open and can visually recognize it, but cannot remember the details needed to retrieve it (URL, file name, search keywords) for task resumption. 
It is worth noting that the current design choice of a video player-like interface may not be ideal for users to locate a screenshot from a lengthy history. 
In addition, the method is most effective when documents are visually distinct, as opposed to relying on textual content. 
These limitations will be discussed to motivate other novel designs for screenshot-based information retrieval methods.} 

To validate the effects of \sys{}, we conducted two in-lab studies. The first user study tested the effects of retrieving a web page using our system compared to conventional approaches; namely, examining the browser history or navigating the web to reopen a page. 
The results indicated that \sys{} decreased the time it took to retrieve a web page and increased productivity by 35.9\% on average, compared to the control condition (without \sys{}). 
In the second study, we validated the software in the context of task resumption, where participants visited the experiment over two sessions with an interval of at least 48 hours between visits. 
The results showed that \sys{} helped participants quickly reopen primary documents, reconstruct their mental context, and retrieve relevant resources used during the first visit. 

The research contributions of this work are as follows: 
\begin{itemize}
  \setlength\itemsep{.1em}
    \item The idea of using a visual history of screenshots to retrieve digital resources, such as software, files, and web pages, that users want to reuse
    \item Exploration of design goals for this method through the development of \sys{}
    \item Quantitative and qualitative evidence that demonstrates the effectiveness of the system in retrieving digital resources using a visual history of a computer screen.
\end{itemize}
The results of this study reveal the design challenges involved in implementing such a method, and they help us understand users' positive and negative perceptions (concerns about privacy and disk space) on self-tracking in computing tasks.  


\section{Envisioned Interaction}

In this section, we present an overview of an envisioned interaction with ScreenTrack in a hypothetical scenario. 

\revise{
Sasha decided to move to a two-bedroom apartment so that she could have an extra bedroom that she can use as both a study room and a guest room occasionally. 
She was debating whether she should buy a sleeper sofa, which looks like a regular sofa. 
She typed ``sleeper sofa'' into a search engine and started surfing the web to find and buy one that she liked.
She browsed tens of products and came across a daybed that she liked because of its contemporary look. 
However, she thought this daybed looked somewhat uncomfortable. 
She decided to keep looking at other kinds of sleeper sofa, eventually finding one that seemed reasonable, if less than perfect. 
She decided to think it over for a few days, since furniture is expensive. 
}

\revise{
Over the weekend, Sasha had a chance to visit a furniture store to see more options. 
She lay down on a few different daybeds and sleeper sofas and realized that a daybed is actually not that bad, given that she usually prefers a firm sleep surface. 
In the worst case, she could switch bedrooms when a guest spends the night at her apartment. 
The daybeds available at the furniture store were not in her price range, so she decided to go for the one online that she had liked. 
That night, she opened a web browser and tried to surf the web to revisit the web page that showed the daybed. 
However, she could recall neither the name of the daybed nor the website where she found it. 
Sasha went through her browsing history and tried searching the web for shopping sites with the keyword ``daybed,'' but the search returned a lengthy list of web pages.
Instead of visiting each web page one by one, she opened \sys{} and opened the screenshot history from the first day of shopping. 
She knew that she had browsed furniture before going to bed, so she started working backwards through her screenshot history by scrubbing the timeline (Figure 1-B). 
She was able to recognize the exact product that she liked and realized she had never heard of the website before then, thinking, `No wonder I couldn't remember where it was.'
She clicked the \textbf{Open} button (Figure 1-D), and the web page opened in her web browser. She added the product to her cart right away, without a second thought. 
}
\section{Related Works}

\sys{} builds upon prior work on: (1) facilitating resumption of interrupted tasks, (2) organizing digital resources by task and time, and (3) using visual histories for mental reconstruction. We discuss prior work in these domains to provide context for the design choices made for \sys{}. 
\subsection{Strategies for Resuming Interrupted Tasks} 

Given the ubiquity of information and communication technologies, external interruptions are inevitable --- email alarms, unexpected encounters in a work space, notifications that prompt users to check a smartphone or other device, and so on~\cite{Czerwinski:2004:DST:985692.985715,Gonzalez:2004:CCM:985692.985707,doi:10.5465/amr.2003.10196791, Iqbal:2007:DRC:1240624.1240730}.
There have been multiple approaches to address this problem: suppressing sources of interruptions~\cite{Dabbish:2004:CIA:1031607.1031638, doi:10.2307/2667031, Kovacs:2018:ROB:3290265.3274364, newport2019digital}, controlling the timing of interruptions~\cite{Adamczyk:2004:EID:985692.985727, Iqbal:2007:UDM:1240624.1240732},  and facilitating resumption of interrupted tasks~\cite{Dragunov:2005:TDE:1040830.1040855_tasktracer,  Jo:2015:EAR:2702123.2702340,5090030, doi:10.1177/154193120605000518, smith2003groupbar}. 
Among these, the idea implemented in \sys{} falls into the last category, since it aims to facilitate task resumption by reducing the time it takes.

Researchers have studied the various cues that knowledge workers use to resume interrupted tasks effectively. Their findings suggest the importance of visual cues in facilitating task resumption. 
Iqbal and Horvitz found that the amount of visual cues (viz., windows that are not obscured) corresponding to a suspended task serve as a reminder to resume the task~\cite{Iqbal:2007:DRC:1240624.1240730}.
Similarly, application windows that are left open can be a cue for programmers to recover the context of a task~\cite{Ko:2005:EDR:1062455.1062492}.
Jo et al. developed gaze-based bookmarking, significantly reducing the time taken to resume reading~\cite{Jo:2015:EAR:2702123.2702340}.

Task resumption has been studied actively in the context of software engineering, as programming is a complex activity that involves multitudes of files and resources. 
Parnin and Rugaber found that programmers rely on the most recent visual state for resuming interrupted programming tasks; programmers tend to run their programs and navigate code to resume tasks~\cite{5090030}.
Researchers investigated the effects of displaying programming activities in chronological order, in terms of both success rate and subjective ratings from users~\cite{Parnin:2010:ECR:1753326.1753342}. 
In particular, the authors evaluated participants' perceptions of the potential of providing an instant replay of screenshots, which is the core idea of \sys{}.
This approach of replaying a code change history has been an active topic in software engineering, as an alternative to coarse-grained version control systems~\cite{5970150,6645254}.  
Existing literature suggests that being able to replay changes in a (reverse) chronological order can be a rich cues for resumption of complex tasks. 

\subsection{Organizing Digital Resources by Task and Time} 

Organizing documents and files has been a challenge for knowledge workers. 
Researchers have developed methods to organize digital resources by task or by activity. 
For example, GroupBar lets users manually organize opened documents by task and makes it possible to restore or minimize batches of documents, which can be useful to quickly switch between tasks~\cite{smith2003groupbar}.
\revise{Taskpos{\'e} visualizes multiple application windows whose proximities are calculated based on usage patterns and transitions between them~\cite{Bernstein:2008:TEF:1449715.1449753}.}
TaskTracer implements a similar approach in an automated fashion whereby the program keeps track of user activity and system events to associate documents with a particular task~\cite{Dragunov:2005:TDE:1040830.1040855_tasktracer}.
While this can be an effective solution to organize resources, users may need to give additional consideration to the concept of a task, and what constitutes a task may vary across contexts. 

In contrast, a number of applications have taken temporal approaches to archiving and retrieving digital resources, diverging from the task-centric approach. 
LifeStreams takes an approach that suggests documenting electronic files temporally, a concept that is prevalent in modern software with familiar features like ``recent documents'' or a browsing history~\cite{Freeman:1996:LSM:381854.381893}.
The idea of time-machine computing is to allow users to go back in time and retrieve files from a previous version of the desktop, similar to the backup system of the same name included in macOS~\cite{Rekimoto:1999:TCT:320719.322582}.
Laevo is a temporal desktop interface that can organize active applications and web pages based on activities laid out temporally, in a manner resembling a Gantt chart~\cite{Jeuris:2014:LTD:2642918.2647391}. 
For example, by clicking a box that represents an activity from the past or which is planned in the future, one can retrieve all of the activity's associated applications and digital resources. 
\revise{YouPivot is yet another system that leverages the temporal proximity of a task to another task and enables contextual searches\cite{Hailpern:2011:YIR:1978942.1979165}.}
\sys{} is similar to these time-centric tools insofar as it leverages humans' episodic memory with rich contextual cues from adjacent tasks and lets users locate a time window of interest. 
However, it also provides rich visual cues to help users reconstruct their mental context and recognize digital resources, whereas the other tools reviewed in this section rely on symbolic information(text, icons).  
\subsection{Using Visual Histories for Mental Reconstruction} 

Many previous studies found potential in using visual histories of activities to facilitate task resumption by helping users reconstruct their working contexts. 
\revise{For example, participants could determine whether they saw a given picture from an extensive collection over a two-day interval with 90\% accuracy~\cite{Standing1970}. 
In other studies, human could spot even a subtle change~\cite{Brady14325} and recall detailed context from an image~\cite{10.1145/1240624.1240636}.
Researchers have made effective use of the capability to recall and recognize visual images in a productive setup.}
One approach to reconstructing a past working context is to provide users with memory cues in the form of visual information in chronological order, such as by writing ``biographies'' with icons~\cite{Lamming94"forget-me-not"intimate}, recording video footage of an office environment~\cite{10.1007/978-1-4471-0105-5_14,lamming1992activity}, archiving a user's web browsing history~\cite{Russell_Retrospective}, or identifying task boundaries in programming environments~\cite{Safer:2007:CES:1321211.1321235}. 
Some systems provide multimodal replay (viz., visual and symbolic), which can not only help users recover context visually, but also retrieve a symbolic state for learning or collaboration~\cite{Grossman:2010:CCE:1866029.1866054, lee2015live, 6645254, Fouse:2011:CSS:1979742.1979706}.
In particular, retrospective cued recall, which uses screen recordings of computing tasks, has been employed as a usability testing method and found to be accurate even with lengthy delays~\cite{10.1007/978-3-642-02806-9_108, Russell2014, Russell_Retrospective}. 
Recently, Rule et al. reconfirmed the benefits of using a visual history for mental reconstruction and enumerated the design implications of such software by exploring various design options (animation vs. still image, screenshot size, crop region) and their effects~\cite{Rule:2017:UVH:3149968.3149977}. 
Beginning from the idea of leveraging the benefits of using a visual history---already confirmed by numerous researchers---we extend it by using a visual history as a method of retrieving digital resources for end users.
\section{\capsys{}: Design goals }

To test the idea of using a visual history of a computer screen to retrieve digital resources, we designed and implemented \sys{}. 
In this section, we discuss the design goals for effective retrieval of digital resources. 
\subsection{Presenting Multimodal Cues in Temporal Order} 
The core idea of \sys{} is to provide a time-lapse video of screenshots that permits users to later open the frontmost (or active) application shown in the screenshot.
\sys{} leverages humans' capability to reconstruct a mental context and recall contextual information from a visual history~\cite{Rule:2017:UVH:3149968.3149977, Russell_Retrospective,Russell2014}. 
Existing methods of recording history stored in different software operate using primarily textual information. 
Some categories of information (e.g., file names, folder structure) must be manually created to be informative, while others are out of users' control (web page titles, URLs).  
As users may find it challenging to recall the actual contents of a document from such abstract information, \sys{} complements existing methods of retrieving digital resources by giving users visual cues in temporal order. 
This means that \sys{} provides \textit{multimodal} --- both visual and textual --- information. It combines screenshots with the textual information --- or \textit{metadata} --- available from existing methods, including file and folder names, URLs, page titles, and application names. 
A series of studies has shown that displaying textual information alongside images can significantly improve both recall of past working contexts and comprehension of the details of previous tasks~\cite{ding1999multimodal, RePEc:bla:jamest:v:46:y:1995:i:5:p:340-347}.

Similarly, the time-lapse video generated by \sys{} distinguishes itself from pure screen recordings in two aspects. First, screen recordings only provide visual cues (implicit), whereas the multimodal cues available in \sys{} can offer users metadata (explicit). Second, a user can open a document from a screenshot in \sys{}, while existing recording software offers no means of recovering resources captured in a video at any level. i.e. one cannot copy text from a screen recording, whereas, in \sys{}, one can open a document from the video and copy text.

Lastly, we made significant use of the findings from Rule's et al. study: we used full-screen thumbnails instead of cropped images, implemented timeline scrubbing, preferred still images over animations, and automated screenshot captures at a fixed interval~\cite{Rule:2017:UVH:3149968.3149977}. 
Given the temporal representation of cues, \sys{} is patterned after a video player. 
\sys{} provides users with three different ways of navigating the visual history of their screens (See Figure~\ref{fig:userinterface}-B): 
users can play the history as a time-lapse video with speed control (Figure~\ref{fig:userinterface}-E); 
scrub through the timeline by clicking and dragging the slider or clicking a spot on the timeline (useful for locating a region of interest); 
and use buttons labeled < and > to display the previous or following captured image. 
Overall, a visual history of screenshots with metadata provides users with rich multimodal cues for them to easily reconstruct mental contexts, and to efficiently retrieve digital resources from the past. 
\subsection{Cross-application Retrieval from a Screenshot}
\sys{} utilizes a visual history of computer screenshots and can inherently document any kind of software that appear on a screen. 
Most modern software provides a way to access a history of previously used files or web pages.
For example, Google Chrome can record the history of web pages accessed by its user over the last (e.g.) 90 days, and Microsoft Word can show a list of recently opened files and documents.
However, each application tracks recent items independently of others, so no application keeps tracks of cross-application usage of a computer and can provide cues for users to understand and utilize the implicit dependencies between files and resources: Google Scholar may always be open while a user is writing an academic paper, a stock photo site may always be open while a user is editing presentation slides, and so on. 
Rule et al. found that users made significant use of ``implicit cues in their working artifacts to reconstruct mental context, particularly for relationships between artifacts''~\cite{Rule:2017:UVH:3149968.3149977}.
As one may need multiple documents and multiple web pages open for a task, the ability to find and recall resources associated with a task is essential~\cite{Hailpern:2011:YIR:1978942.1979165}. 
As \sys{} present screenshots in chronological order and users interleave multiple applications within a task, they can retrieve resources that are relevant to their primary documents by navigating the timeline.

While a captured image of a computer screen inherently works for any kind of application that appears on a computer, capturing metadata is not trivial. Each application has different kinds of metadata meaningful to users, and it may not be readily available to other software. 
For example, in the case of a web browser, the URL and title of a web page are relevant to users, while in the case of a word processor, a document's name and location on disk are relevant to users.
Certain applications offer little or no metadata (e.g., Dictionary, Chess), and certain applications have the property of working with groups of files as one unit, as is the case when opening a project in an integrated development environment, or when opening a library of entries in a bibliography software. 
To make \sys{} work across applications, we categorized frequently used software (office/productivity suites, programming environments, web browsers) and designed scripts to capture metadata differently for each. 
In addition, labels shown in the \sys{} user interface are adapted to the category of the frontmost application for the current screenshot. For instance, if the active application is a web browser, the item label reads ``Page URL'' (as seen in Figure~\ref{fig:userinterface}-C). If the active application is Microsoft Word, the item label will instead read ``File Directory'', and the button next to the label may be activated to open the enclosing folder. 
The current version of \sys{} provides four different categories and covers most frequently used software. 
The list of categories and applications is detached from the main algorithm so that it can be easily updated from a central repository of application-category mapping. 
In summary, metadata  can be captured across applications in \sys{}.

\subsection{Addressing Users' Privacy Concerns}

Even though more people are comfortable with recording their personal data in the emerging culture of self-tracking and the quantified self~\cite{Choe:2014:UQP:2556288.2557372, Epstein:2015:NCW:2675133.2675135, Neff:2016:SEL:3056035, kopp1988self}, there remain critical concerns that self-tracking technologies may put one's privacy in danger~\cite{Barkuus03location-basedservices,Lupton2016,Raento:2008:DPS:1410427.1410436}. 
We designed \sys{} to address the potential concerns involved in recording a computer screen, given that personal information (such as a credit card number) may be displayed on-screen, as anticipated in the previous study~\cite{Rule:2017:UVH:3149968.3149977}. 
Firstly, \sys{} is designed to use only local storage (instead of cloud-based storage). Therefore, cross-device functionality was not considered (such as stitching videos from multiple computers), since these features would require captured images to be transmitted from one device to another. 
\sys{} is meant for use on a single computer. 
In addition, all the images captured in \sys{} are designed to be accessed only through the software, not through external means. 
Only those explicitly granted access to the computer can review the time-lapse video, assuming the computer is password-protected. 
In addition, it is trivial to incorporate them the software in ways that conform to modern security standards --- giving users the option to protect the software with a password or biometric authentication and using encryption to deny third parties access to captured images. 
Currently, we plan to implement these features for a future long-term deployment study and the eventual release of the software. 
Lastly, the time-lapse video is designed to be ephemeral. Storing only recent data (recent $N$ days) may alleviate some users' concerns about privacy~\cite{doi:10.1080/15710880600608230, Xu:2016:AAV:2818048.2819948}. 


\begin{figure}
\centering
  \includegraphics[width=.95\columnwidth]{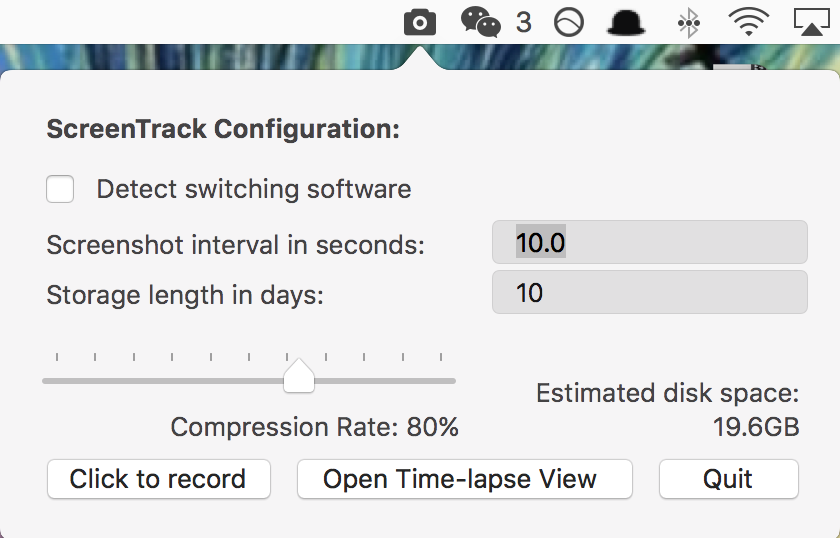}
  \caption{\sys{} configuration: Users can configure the recording parameters and control how much disk space is needed. Clicking Open Time-lapse Video will open the main interface shown in Figure~\ref{fig:userinterface}.}
  ~\label{fig:configuration}
  \vspace{-20pt}
\end{figure}

\subsection{Moderate Disk Space Usage}

\sys{} is designed to use a minimal amount of disk space, given the effectiveness of a visual history in mental context reconstruction and retrieval of digital resources.
In theory, as always-on screen recording software, captured screenshots could easily occupy all available disk space within a few months. 
Importantly, making the time-lapse video ephemeral and limiting the period for which \sys{} will keep a time-lapse video are essential in limiting the amount of disk space used. 
This approach is common with other kinds of recording software and equipment (e.g., dashcams, security cameras, action cams).
Users can also configure the time-lapse video quality (image compression and capture interval) to decrease the amount of disk space that the screenshot repository will require. 
However, considering the trade-off between the quality of the captured visual history and the amount of space needed, \sys{} imposes minimal values for some parameters (such as a minimum dimension of 320 px and a maximum capture interval of 60 seconds) based on the findings from a previous study~\cite{Rule:2017:UVH:3149968.3149977}.

In addition, \sys{} implements a few algorithms that can reduce the total amount of disk space needed. 
For example, Figure~\ref{fig:configuration} shows how one can choose to capture a screenshot whenever the active application is switched, so that one can choose to specify a longer interval without missing the opportunity to capture each application used, similar to the motion-activated security camera. 
In addition, \sys{} detect if the newly captured image is worth storing. 
If two consecutive images are identical, it stores only one image, given that the following image does not add any new information. The single image is instead reused for several frames of the generated time-lapse.  
Using the capture configuration, \sys{} provides a conservative estimate of the amount of disk space required, given a users' input on the length of video storage(See Figure~\ref{fig:configuration}). 
For instance, based on current practice, approximately 20 GB are needed to store 80 hours' ($\approx$ 10 days') worth of data, given a reasonable configuration: screenshots every 10 seconds, 80\% compression (assuming a 1440 $\times$ 1080 capture resolution).  
Informing users of approximate disk space requirements enables them to reconfigure the software to strike a balance between capture length and resolution (i.e., both time and size) of the replay, and to thus control the disk storage space they would like to dedicate for \sys{}. 
\subsection{Implementation} 

We designed and implemented \sys{} in Swift for MacOS devices. 
The program also uses AppleScript and the Accessibility API to capture metadata, launch software, and open documents and websites. 

\sys{} saves two kinds of data in the recording process: captured screenshots and corresponding metadata.
Each screenshot is stored as a JPG image, and all metadata is stored in a JSON file with time stamps and paths to the corresponding images.
During use, \sys{} remains present in the menu bar as a camera icon.
Clicking the icon displays a configuration window as seen in Figure~\ref{fig:configuration}.
Users can specify the screenshot interval and \revise{per-image} compression ratio before starting recording in the interface.
Collected data is stitched or segmented by day. Users can access a time-lapse video by selecting a date (See Figure~\ref{fig:userinterface}-F).

\section{User Study Result}

We conducted two in-lab studies to validate the effects of \sys{} in two different settings. 
Each studies investigate the following research questions (RQ1) Does using \sys{} reduce the time it takes to retrieve a digital resource from the past compared to existing methods? (RQ2) How does \sys{} support users in the context of resuming suspended task? (RQ3) What are users' perceived values and potential concerns of \sys{}? Each question will be explored in each section. 
\subsection{STUDY 1: Revisiting a Web Page of Interest}

\subsubsection{Method: Recruitment and Study Procedure}

We were interested to see if using \sys{} would cause a reduction in the retrieval time. 
This could lead to a productive gain overall in computing tasks by complementing existing methods, when there is no single, direct way (e.g., a bookmarked web page or a file on the desktop) to retrieve a necessary resource. Modern knowledge workers often encounter this situation.  
Study 1 is designed to investigate how \sys{} helps users retrieve a previously opened digital resource (a web page, in this case) from a screenshot. 
To that end, we simply measured the retrieval time, which is the time it takes users to retrieve a web page that they recall.

\begin{figure} 
\centering
  \includegraphics[width=\columnwidth]{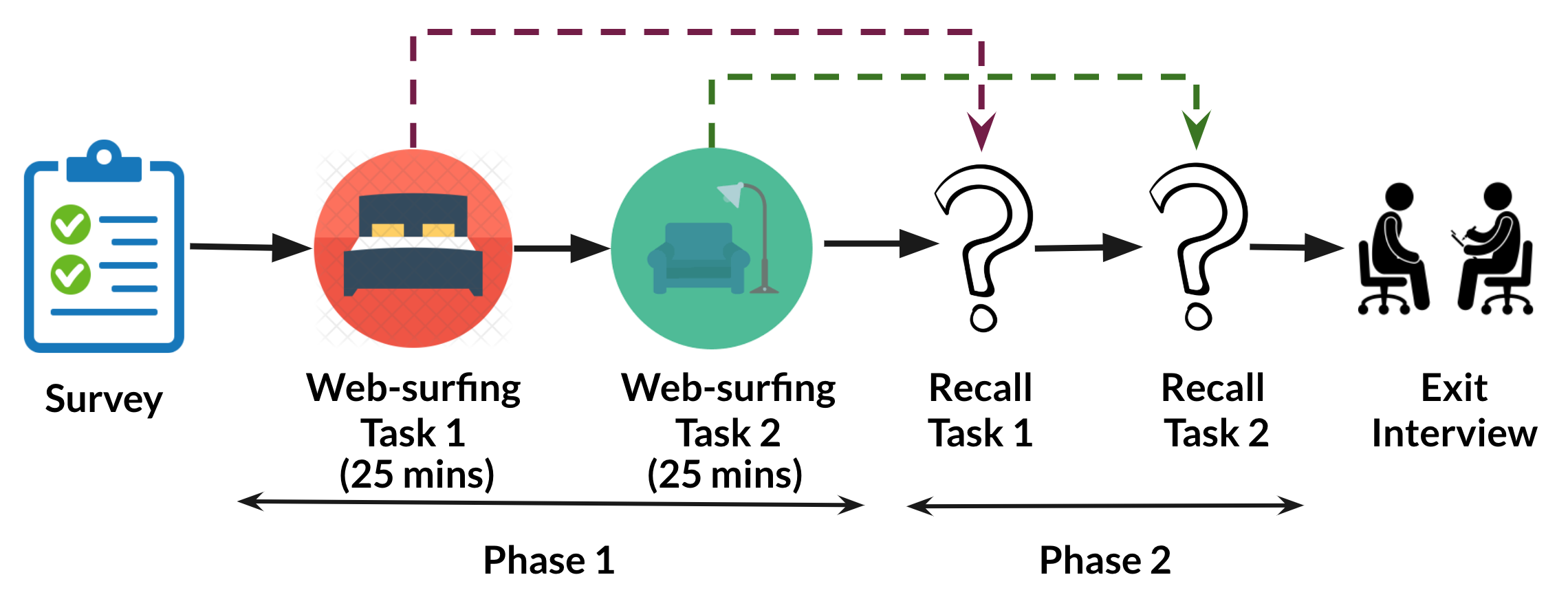}
  \caption{\sys{} The user study procedure. The user study was composed of seven parts: (1) consent form, (2) survey, (3) web-surfing task 1, (4) web-surfing task 2, (5) recall task 1, (6) recall task 2, and (7) exit interview. }
  ~\label{fig:Study1}
  \vspace{-15pt}
\end{figure}

We recruited 20 university students (7 of whom were female) through a student mailing list at the authors' university. 
The average age of the subjects was 26.2, ranging from 22 to 35. 
The results of the demographic survey indicated that all the subjects self-assessed their digital literacy as `Average', `Above Average', or `Excellent', and they all frequently surf the web (at least once a day). 
We invited the subjects to a lab and provided them a laptop on which \sys{} was installed.
Upon arrival, all subjects signed a consent form and filled out a demographic survey (See Figure~\ref{fig:Study1}). 

In Study 1, subjects performed an online shopping task. We asked them to browse furniture products (such as couches, media stands, and coffee tables) for two different room types (a living room and a bedroom) until they had decided which products to purchase per type. We asked subjects to select at least three products per room within 25 minutes. 
The task was divided into two phases: (1) the web-surfing phase, in which subjects were asked to browse online shopping websites; and (2) the retrieval phase, in which a study moderator asked each subject for their favorite products per type and instructed them to retrieve the web page on which the chosen product is listed. 
Each subject did both tasks under two different conditions: the control condition and the \sys{} condition.

\noindent\textit{(Phase 1)}
In Phase 1, we asked subjects to pick three furniture products that they would like to purchase. Subjects were told to assume that they were moving into an empty version of the home that they were currently living in. 
We split rooms across conditions for each subject; in one condition, we asked them to browse furniture for their living room; in the other, we asked them to browse furniture for their bedroom. 
The purpose of this split was to avoid having subjects pick the same set of furniture under two different conditions; additionally, there was a risk that subjects would run out of ideas if they were asked to choose furniture for the same room twice.
Four combinations of room and condition were thus created: C1 (control-Living room), C2 (\sys{}-Bedroom), C3 (control-Bedroom), and C4 (\sys{}-Living room).
Each subject was assigned one of two pairs, either (C1, C2) or (C3, C4). 
We counterbalanced the number of subjects per sequence of pairs, for a total of 5 subjects in each of four different sequences: (C1-C2),(C2-C1),(C3-C4),(C4-C3).

\noindent\textit{(Phase 2)}
Before Phase 2, everyone was introduced to \sys{} and went through a short, interactive tutorial on how to use \sys{}. 
We made sure that each subject knew how to open a document and a web page from a screenshot using the Open button and had tested each way of navigating through time (playing the time-lapse, playing it at a different speed, scrubbing through the timeline, using the buttons to show the next or previous screenshot). 
After Phase 1, all subjects were asked to answer the following questions per product category (for a total of three products): (1) Could you describe the details of your favorite product (e.g., couch)? (2) Could you open the web page that contains the product you just described?
The first question was chosen to let subjects pick a single furniture product among the products they browsed. 
The second question was chosen to measure the time it took to retrieve the web page. 
For the second question, if they used \sys{} in Phase 1 for the product in question, we asked them to find the web page using the time-lapse video in \sys{}. 
In Study 1, we configured \sys{} with a screenshot interval of 5 seconds and a screenshot resolution equal to 40\% of the screen's resolution.
For products found when \sys{} was not in use, subjects were free to use any conventional method, such as typing a URL, exploring the browser history, searching for keywords on any web page, and/or navigating to the web page by clicking hyperlinks and menus.
In both conditions, we measured the time between the very first interaction with the computer --- typically, launching a web browser in the control condition or launching \sys{} in the experimental condition --- and the time of the user's interaction that opened the target web page. 
We repeated the same questions for the second task completed in Phase 1. 
With 20 subjects, we collected 60 data points per condition (3 products per task, 20 subjects), making for a total of 120 data points with two different conditions. We conducted an exit interview to ask subjects general questions about the system, which we will present the result in the later section. 
We recorded audio and the screen of the laptop during the entire study session, from the point when subjects finished the survey until the end of the exit interview. 
We manually annotated how each subject retrieved the target web page, as well as how long it took. 
We double-checked that the page they retrieved actually contained the product described and confirmed that they visited the page during Phase 1 from the screen recording. 

\subsubsection{\textit{STUDY 1 Result: Time Reduction in Resource Retrieval}}

The result indicated that subjects can retrieve a target web page quicker with \sys{} than without it. 
Under the \sys{} condition, the average time to retrieve a web page of interest is 27.1 seconds ($\sigma = 26.3$) compared to 42.9 seconds ($\sigma = 30.8$) under the control condition. This is equivalent to a 35.9\% gain in productivity. 

We ran a repeated-measures ANOVA to test the significance of the effects of the following variables on retrieval times: $Condition$ (control or \sys{}), $Trial$ (the first, second, or third product chosen in each task, to test for a within-condition learning effect), $Room$ (Bedroom or Living room), and $Order$ (the first or second task, to test for a learning effect between tasks). 
Because distribution of the retrieval times does not satisfy the normality based on Shapiro-Wilk test ($p = 1.103 \times {10^{-11}}$), we performed log-transformation of all data points and confirmed the normality of the result (Shapiro-Wilk test, $p = 0.5936$).
The test result showed that the difference between the mean retrieval times for each value of $Condition$($F[1,96] = 27.3, p < .0001$) was statistically significant.
This result indicates that using \sys{} reduced the time taken to retrieve a previously visited web site. 
In addition, mean retrieval times according to other groupings were also statistically significant: $Trial$($F[2,96] = 9.05, p < .0001)$), and $Room$ ($F[1,96] =  9.6467, p < .01$), indicating that subjects retrieved living room products more quickly on average. 
Meanwhile, $Order$($F[1,96] = 5.7528, p <. 05$) indicated that subjects completed the second task more quickly on average. 
We figured out that there is no significant interaction effect among the factors of our interest (all of the p-values > 0.05).

\begin{figure}
\centering
  \includegraphics[width=\columnwidth]{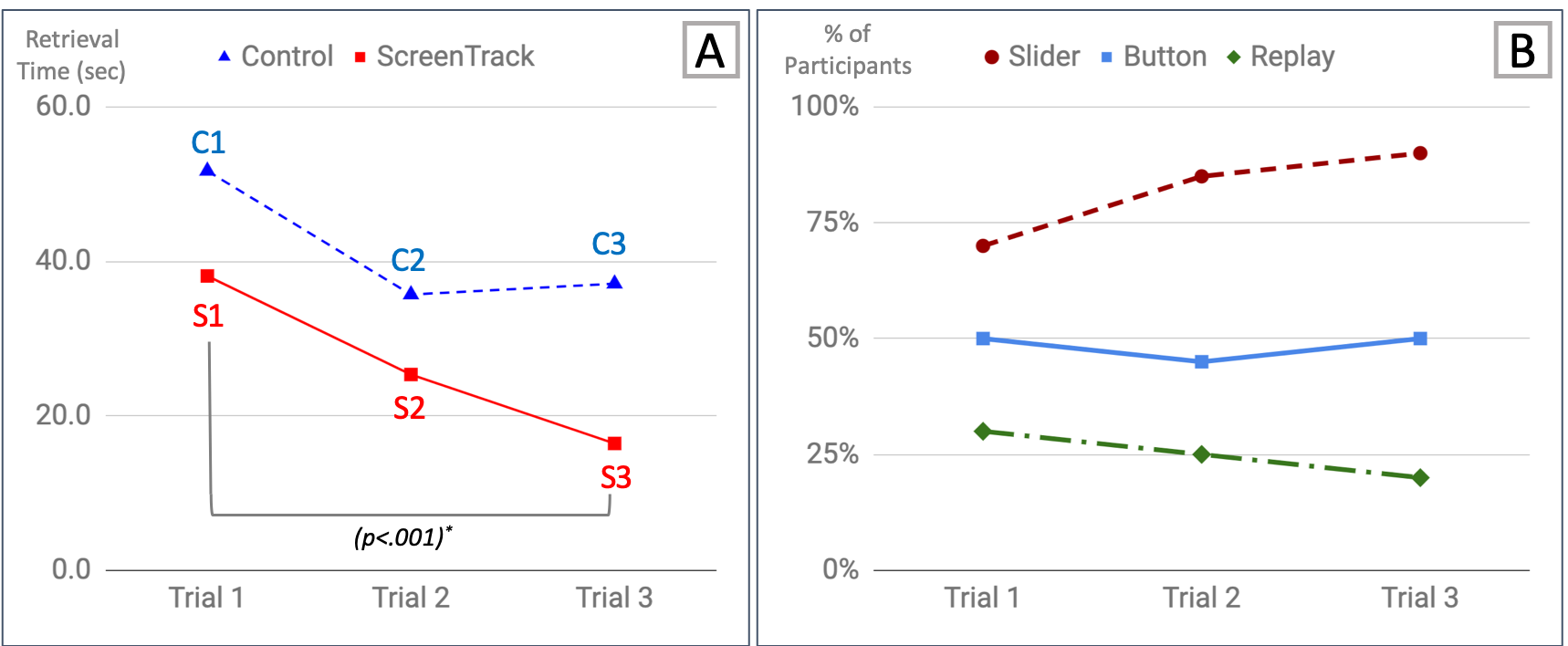}
  \caption{The effects of Trial: (A) Average retrieval time decreased over trials, especially in \sys{}. (B)  As the trials progressed, more subjects used the scrubbing method to browse the captured screenshots, which may account for the decrease in retrieval time in (A). 
  }~\label{fig:Trial}
  \vspace{-20pt}
\end{figure}

We further analyzed the data to understand the within-condition learning effect (\textit{Trial}). 
This is to investigate whether consistent use of \sys{} could lead to greater benefits to productivity. 
Figure~\ref{fig:Trial}-A shows the average retrieval time per trial and confirms that subjects could more quickly retrieve a web page in later trials, especially under the \sys{} condition.
We ran Tukey HSD test to pairwise compare the retrieval times in order to see the significance of the difference between the times across the trials. 
This revealed a statistically significant difference between the first trial (S1) and the third trial (S3) under the \sys{} condition ($p < .001$). A marginally significant difference between (C1-C3) and (S2-S3) was also found ($p < .1$). 
Based on the trend shown in Figure 4-A and the result of Tukey HSD test, we can conclude that constantly using ScreenTrack has capability to improve users' performance. 
We reviewed the video recordings and found two potential reasons that may explain this trend beyond the learning effect. 

Firstly, we observed those who simply replayed the time-lapse video from the beginning took significantly longer to retrieve their page of interest than those who navigated the video using either the slider or the frame-by-frame < and > buttons. 
Some subjects scrub through the timeline until they find their product of interest, and may then press the < and > buttons to finely adjust the timing. 
This approach took less time than the approach in which subjects watched the replay from the beginning. 
To see the usage patterns over trials, we annotated and calculated percentages of control usage (replay, slider, buttons) across subjects. 
In general, note that the majority of people used the slider rather than replaying the time-lapse video. 
We can see that on average, the subjects used the replay function less and used the slider (timeline scrubbing) more in later trials, as seen in Figure~\ref{fig:Trial}-B. 
While it could not be confirmed statistically, increased slider usage as subjects became more familiar with \sys{} might explain why subjects became quicker at retrieving a web page. 

Secondly, the way that \sys{} works gives users additional information as they use it. 
For example, when a user opens \sys{}, the slider will be positioned at the most recent screenshot relative to the selected date. 
One common pattern for using the horizontal slider (timeline) is to grab the handle and drag (or scrub) it to the left. 
As they drag the handle, the video view is updated to respond to the handle motion. 
Therefore, subjects were shown other products that they browsed before while they were trying to locate the product. 
Even though they did not know exactly where their products of choice were located on the timeline for Trials 2 and 3, subjects should have known approximately where other \textit{types} of products (e.g., coffee tables) were on the timeline. 
In that sense, we strongly believe that this could be another reason for the continued improvement in retrieval times. 
The result of Study 1 demonstrated the potential benefits of using a visual history of a computer screen for the purpose of retrieving web pages. 


\subsubsection{Interview Result: Perceived Values and Concerns}

In this section, we briefly summarize the result of thematic analysis we have done on the exit interview of Study 1. 
We asked four general questions (benefits, concerns, willingness to use in practice, open-ended suggestions) to learn their perception on the system and potential benefits and concerns that they found. 


In the interview, 15 subjects (out of 20) said that the user interface of ScreenTrack is easy to use and intuitive. 
Subjects were favorable to the interface because it is similar to watching a video.  
A number of people commented about the benefits of multimodal cues and the feature that they can retrieve digital resources. Pr's comment nicely summarize the benefits:

\begin{displayquote}\small{
P4: "It is easier to use compared to the web browser history where I have to actively search for the keyword. It follows a chronological order that can even further help me back to chase what I did and which web page I clicked. of course, it can direct me right to the web page which is pretty cool and easy."}
\end{displayquote}

Half of the subjects (10/20) discovered the benefits of retrieving digital resources across application from a time-lapse video. 
They also agreed that the visual history can help users recall more details about previous working context. 
The following comments well resonates with the challenges of modern knowledge workers and describes how \sys{} address such a challenge and potentially enhance productivity.

\begin{displayquote}\small{
P5: "Because I will write many papers, there are many files you know. Sometimes I forgot to rename them. So it is really difficult for me to find the exact files in a short time. 
(...) So with this tool, I think I can save lots of time to select, to find the files I want to even across many different applications."
}\end{displayquote}


Subjects raised a few concerns on using the \sys{} during the interview. 
Subjects worried that this software would record personal information and might leak this data to someone else, like credit card information or home address.

\begin{displayquote}\small{
P16: "I mean it is recording me, so that's a concern,
(....) if it goes into somewhere then it might be in trouble, that's a problem."
}\end{displayquote}

Lastly, in addition to the common concerns on the disk-space usage, quite a few number of subjects also expressed their hesitance in installing a new software to their computers.
While these concerns were anticipated and considered from the design stage, we realize that there may be actual barriers of using \sys{} in practice.
Features that can enhance awareness and accountability would be necessary to alleviate their concerns on privacy and computational resources. 

\subsection{STUDY 2: Facilitating Task Resumption}

Study 2 is designed to investigate the potential benefits of \sys{} in more complex setups and over longer periods of time. 
The context of Study 2 simulates a scenario in which a user needs to resume a suspended task, like the example scenario given in the Envisioned Interaction.
In summary, we asked subjects to create a deck of presentation slides over two sessions, with a minimum 48-hour interval between sessions. 
With this task, we were interested to see how subjects would retrieve the primary document and associated digital resources when resuming the suspended task.

\begin{figure}
\centering
  \includegraphics[width=\columnwidth]{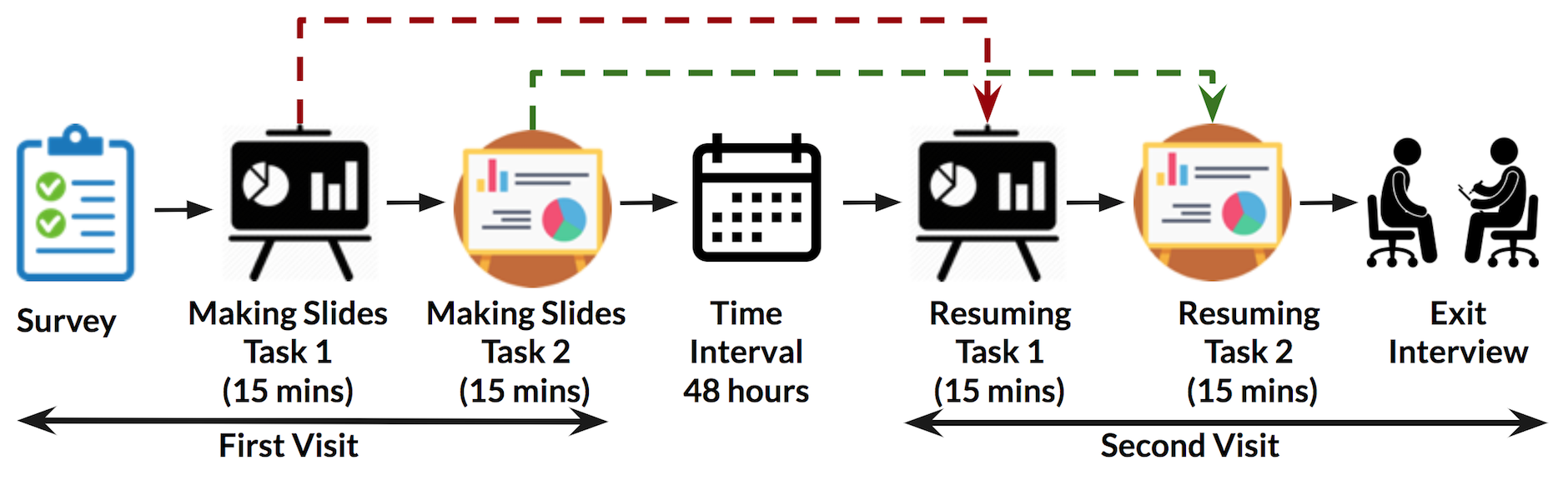}
  \caption{\sys{} The procedure of STUDY 2.The user study 2 is composed of 7 parts: (1) consent form, (2) making slides task 1, (3) making slides task 2, (4) time interval, (5) resuming task 1, (6) resuming task 2, (7) exit interview}~\label{fig:study2}
  \vspace{-20pt}
\end{figure}
\subsubsection{Method: Recruitment and Study Procedure}

For Study 2, we recruited 8 subjects from the author's university.
The average age of the subjects was 25.1, ranging from 23 to 27. 
We asked subjects if they were familiar with presentation slide authoring software (e.g., Microsoft PowerPoint), and all subjects assessed their skill in such software as either ``intermediate'' or ``advanced''. 

Subjects in this study were asked to create two decks of presentation slides under two different conditions: control and \sys{}.
One slide deck was to introduce Hawaii as a place for a family of four to travel, while the other was to introduce Los Angeles as a destination for subjects and their friends to visit. 
These topics were selected for their general relatability through a pilot study. 
The same laptop used in Study 1 was used for this study. 

Study 2 was conducted over two sessions, simulating a situation in which a user has been interrupted and must resume the suspended task after a fixed interval of at least 48 hours.
In their first visit, subjects were asked to perform one task for 15 minutes, and then stop the task. 
After stopping the first task, they began the second task, which was also interrupted after 15 minutes regardless of what they were doing. 
For each subject, one of the two tasks was recorded by \sys{}, which was notified to the subjects.
In this study, there were two factors: Condition (control or \sys{}) and Place (Hawaii or Los Angeles). 
Each subject performed one of two pairs: either (control-Hawaii, \sys{}-Los Angeles), or (control-Los Angeles, \sys{}-Hawaii), in a randomly assigned order.  
The interruption was patterned after a typical external interruption, in which users are stopped suddenly and do not have time to wrap up their work properly~\cite{coraggio1990deleterious,Mark:2005:NTL:1054972.1055017}.
In the beginning of the first visit, they were asked to save the slide deck in a designated folder that the study moderator provided. 
While subjects were not familiar with the folder structure, it was nonetheless well organized and named, such that subjects could navigate it and save their files easily. 
For example, a slide deck for Los Angeles was saved to this folder: $/Documents/Personal/Travel/2019/LosAngelesTrip/$.
The folder structure also contained a variety of other folders that served no real purpose, to improve the realism of the scenario. 
Lastly, we asked subjects to log into the web browser (Chrome) with their own account (i.e., a Google account) before beginning the task.
This way, any browsing history made on the laptop would be logged to the subject's personal account and later synchronized during the second visit. 
This serves to create an environment similar to that of using one's personal computer; otherwise, all subjects' browsing history would appear in the browser.

Each subject's second visit was scheduled 48 hours after the first visit with the exception of one subject, whose second visit was scheduled 70 hours later than the first visit due to a scheduling conflict. 
The 48-hour interval was selected because it is known that within two days, one will forget roughly 70\% of what one has learned without any reinforcement~\cite{10.1371/journal.pone.0120644}. 
Upon arrival at the second visit, subjects were asked to resume the suspended task --- completing the slide decks they created, in the same order as they began them during the first visit. 
For their control condition task, users did not have \sys{}, but were allowed to use any other functionality (e.g., file search, recent documents, file manager [the Finder on MacOS], browser history).
For their \sys{} condition task, subjects were encouraged to use use any tools they desired, including \sys{}. 
The same tutorial (from Study 1) was repeated before running the \sys{} condition task. 
Each task was stopped within 15 minutes of its start time.
Upon completion, we conducted the exit interview as in Study 1.
We recorded audio and the screen of the laptop during the entire study session, from the point when subjects finished the survey until the end of the study. 
Later we manually annotated the screen recording video for analysis
While we were also able to confirm the time reduction benefit in Study 1, we focused on what kinds of natural behaviors would emerge as subjects resumed the suspended tasks with \sys{}. 

\begin{figure}
\centering
  \includegraphics[width=\columnwidth]{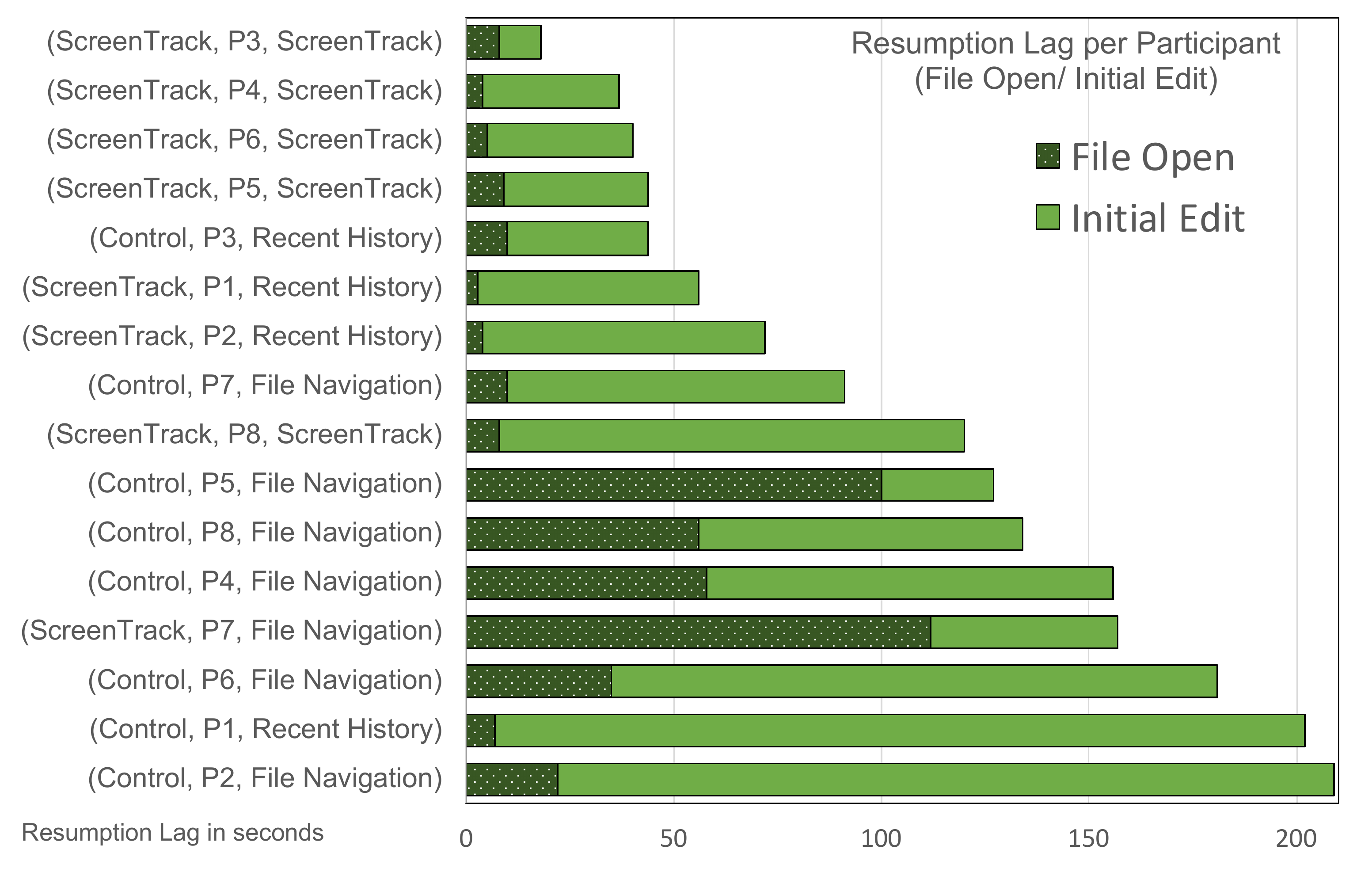}
  \caption{Resumption lag per subject. The label indicates (Condition, Subject ID, The method of file open). The left bars (dark) present the time it took for a subject to open the file. The right bars (light) present the time it took to make the first edit after opening the file. }~\label{fig:resumptionlag}
  \vspace{-15pt}
\end{figure}
\subsubsection{Result: \sys{} Usage in Longer Time Scale.}

The result suggests that using \sys{} is useful in reducing resumption lag overall. 
Resumption lag is defined as the time needed to resume the primary task, and is often determined by the first edit that a user makes to the primary document~\cite{Salvucci:2010:RTC:1753326.1753341}.
Figure~\ref{fig:resumptionlag} shows resumption lag per subject. 
We broke down resumption lag to additionally measure when and how the file was opened (the left side of the bar --- dark dotted bar). 
For each bar, each label is composed of condition, subject ID, and the method that they use to open the file.
Note that not all subjects (viz., P1, P2, P7) under the \sys{} condition used \sys{} to find their presentations file and open them from a screenshot; some of them, even after spotting the file in a screenshot by scrubbing through the timeline, habitually either clicked the application icon or opened the file manager to locate the file. 
We can observe that opening a file from \sys{} did help subjects quickly retrieve the file from a snapshot. 
Those who used ``recent documents''-type functionality were similarly able to quickly open the file.

Once they opened their files, it took a while for subjects to make their first edit of the session.
This is the time dedicated to mentally reconstructing the task context.
As shown in Figure~\ref{fig:resumptionlag}, overall resumption lag in case they used \sys{} is shorter than the ones in control condition with one exception (P8).
We checked the screen recordings for all subjects in \sys{} condition, and all except one subject used \sys{} to review what they did, typically after reviewing their slides. 
While reviewing the slides did provide subjects with the most recent version of the artifact, it did not inform them of the nature of their most recent change was or what they were about to do. 
Watching the temporal history of screenshots might have helped reconstruct their mental context, which can potentially be a reason for the productive gain. 


\sys{} provided richer information than the file opened. 
The visual histories always included the web pages that they used at the first visit. 
For example, a subject was interrupted when they were entering "family friendly restaurants in Honolulu".
The search keyword that the subject was able to see would have provided richer contextual information (e.g., a trip for family of four) about the task than the empty slide that reads "Top places to eat".  
This kind of additional cues cannot be obtained from the primary document itself and would be the benefit only comes from cross-application cues available in \sys{}. 

\begin{figure}
\centering
  \includegraphics[width=\columnwidth]{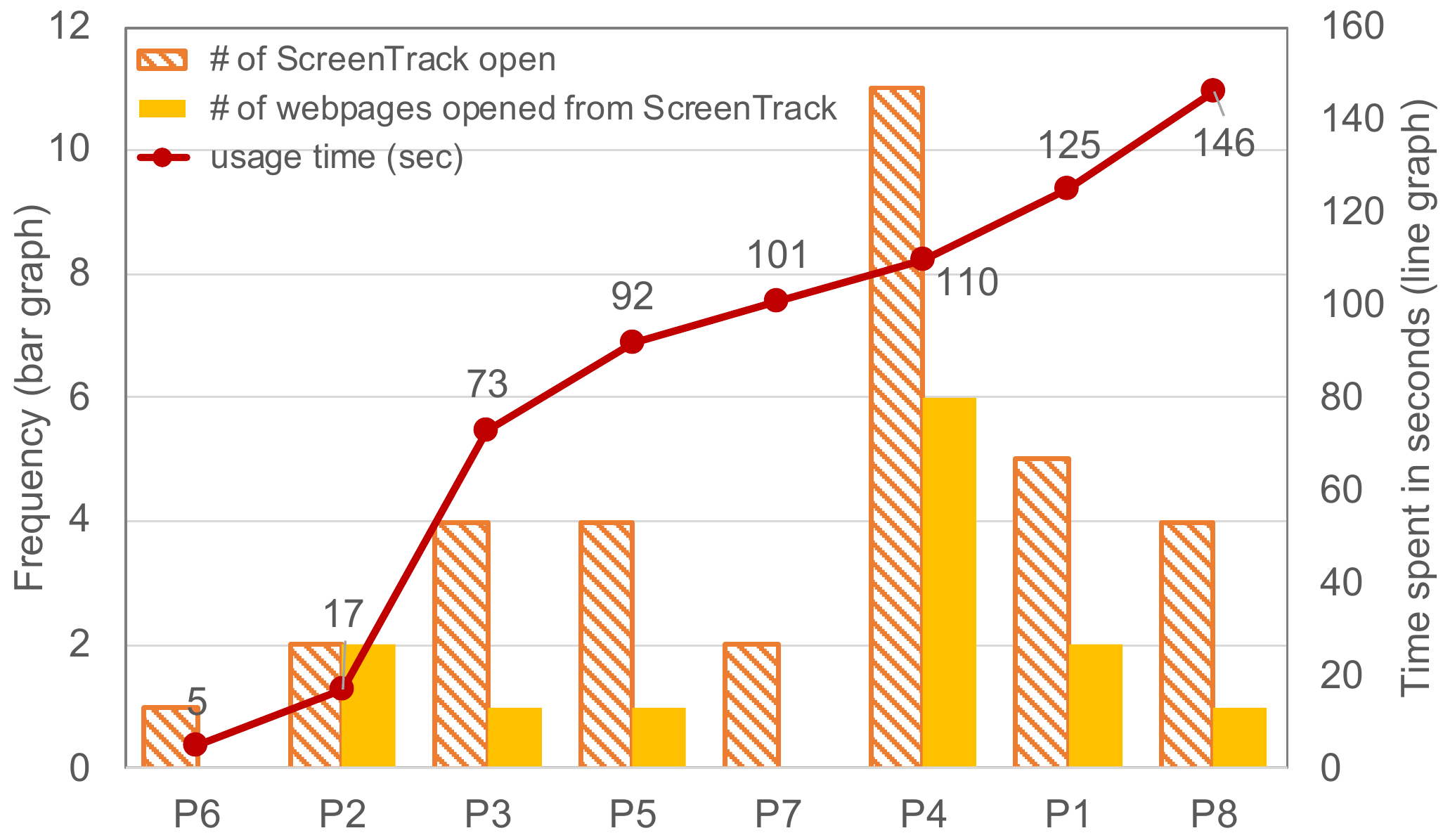}
  \caption{The usage of \sys{} per subject (sorted by the elapsed time) (Left Bar) indicates the number of times that subject used \sys{}, (Right Bar) indicates the number web pages open from \sys{}, (Line) The time(seconds) that subjects spent using \sys{} during the 15 minute-long task at the second visit}~\label{fig:usage}
  \vspace{-10pt}
\end{figure}
We also investigated how the usage of \sys{} vary across the subjects. 
From annotating screen recordings, we counted 1) the number of times that they used ($mu$ = 4.1, $sigma$ = 3.1),  \sys{}, 2) the number of web pages opened from \sys{} ($mu$ = 1.6 , $sigma$ = 1.92), and 3) the time that they spent in \sys{} ($mu$ = 83.6 sec, $sigma$ = 49.9). 
Each subject's results are displayed in Figure~\ref{fig:usage}.
The usage of \sys{} varied much across the subjects; i.e. P6 did not open any web page and almost did not use the application (5 seconds for initial file open), while P4 actively used the application to open six web pages that they used from the 1st visit, not just initially but throughout the session. 
On the other hand, P8 spent the most time (146 seconds) watching the visual history from \sys{} --- 1/6 of the total time spent at the second visit. 
At the first visit, the subject spent most of the times searching information, rather than making slides. 
By the time that the subject was asked to stop the task, the slides was mostly empty and the works she has done during 15 minutes did not translate to the artifact. 
Therefore, the subject started with sparse slides and decided to watch the replay from the beginning to remind themselves what they did last time. 
The observation that some subjects spending time on watching replay throughout the session, and retrieving digital resources from the past shows the limitation of resumption lag as a representative metric to evaluate the effectiveness of task resumption. 
It is because resumption lag only highlights only the initial part of the challenge~\cite{Salvucci:2010:RTC:1753326.1753341}. 
The variance of usage highlights the need of long-term deployment studies as a future work in order to understand how \sys{} can be integrated in their day-to-day computer works. 


\section{Limitation and Future Works}
\revise{One limitation of \sys{} is that it does not scale well over the temporal dimension, for medium- or long-term retrieval.
The current design of \sys{} --- a media player-like interface --- makes it difficult for users to retrieve a document from a lengthy history. 
For example, if a user cannot recall the exact date of an incident, it would be very time-consuming to go through multiple time-lapse videos to identify a snapshot. 
Even within a single video, locating a snapshot without an idea of its position on the timeline is as challenging as navigation of video streams to find information just by visual inspection. 
This timeline design limits the effectiveness of ScreenTrack to scenarios that involve resuming recent tasks, or cases in which users know approximately when a document was used within the last \textit{N} days (e.g., finding a product in the evening after viewing it in the morning, or locating a website visited last weekend). 
Instead, \sys{} relies on a person's episodic memory, which is the ability to remember events based on visual images and their temporal order~\cite{CONWAY20092305}. 
Existing methods do not leverage a user's episodic memory to retrieve resources, as they do not use images sorted in chronological order. In that regard, we believe using \sys{} has value in providing a new retrieval method different from and complementary to existing ones. 
However, it is worth noting that other types of design choices (e.g., query-based~\cite{atlas2016}, sampled or curated snapshots) may be more effective for longer-term scenarios.
This limitation motivates our future works, in which we plan to explore other design alternatives for screenshot-based task organization.
In particular, we plan to develop a new visual representation of screenshots geared towards long-term usage of screenshot-based methods for document retrieval, mental reconstruction, self-reflection, and behavioral change in the context of productivity. 
One design choice we plan to explore is a collage of screenshots that an intelligent system can generate from \sys{} data. 
Such visualizations may be able to serve as representative summaries of what a user did in the past, and may also project how productive the user was~\cite{Christel:2002:CDS:641007.641120}. }

\revise{Another barrier to extending the benefits of \sys{} is the limited ecological validity of the user study. 
While we believe evaluating the system in a controlled setting was inevitable to provide quantitative evidence, given the limited time spent in the lab, we may discover more benefits and challenges associated with \sys{} if tested in practice. 
For example, in real-world use, users would need to go over much longer videos, full of irrelevant information as described previously, which may slow down the retrieval process.} 
We plan to run a long-term deployment study \sys{} and are in the process of developing practical features to address participants' concerns (disk space, multiple monitors, encryption, password protection, tolerance to file path changes).  
\revise{
 Through the extended study, we plan to further explore and understand how users' perceptions change in the long term. Specifically, we wish to study how \sys{} causes behavioral changes, to discover unexpected (both positive and negative) implications of self-tracking in the context of productivity.
 }
For example, if \sys{} can be seamlessly integrated into users' computing activities, allowing them to retrieve documents, web sites, and applications, this may change the perceived value a cluttered workspace has to a user as one motivation to leave documents open is not to lose them. 
Users who currently leave many applications running at once and open dozens of browser tabs may feel more comfortable closing unused resources. 

\revise{Lastly, we note that the suggested method heavily relies on the visual nature of applications and websites. 
The current user study only involves computing tasks that are designed around visual tasks (online shopping, authoring presentation slides) rather than text-heavy tasks (writing in a word processor).
Therefore, the effects of recognizing resources from a screenshot in \sys{} may not translate well to cases where digital resources in a time-lapse video are not visually distinct, as when switching between text documents. 
In the future long-term study, we plan to investigate the extent to which computing tasks that include non-visual resources limit the effects of our screenshot-based retrieval method.
} 

\section{Conclusion}
In this study, we explored the idea of using a visual history for retrieving digital resources such as applications, web pages, and documents. 
To test the idea, we designed and developed \sys{}, and validated it in a series of user studies that explores two different time scales and setups. 
The result showed that using \sys{} led to a productivity gain of 35.9\%. 
We also confirmed that the rich contextual information embedded in the time-lapse videos helps users reconstruct their mental context and facilitates task resumption while moderating potential concerns related to privacy and disk space usage.

\section{Acknowledgements}
We thank all the reviewers for their thorough feedback. We are also grateful to all participants and Edward Powell for their contributions in the user studies; and Prof. Scott McCrickard and Prof. Denis Gracanin for their advises on this work.  

\balance{}

\bibliographystyle{SIGCHI-Reference-Format}
\bibliography{proceedings}

\end{document}